\input harvmac
\input epsf
\noblackbox

% -------------------------AK Definitions
\font\cmss=cmss10 \font\cmsss=cmss10 at 7pt
 \def\inbar{\,\vrule height1.5ex width.4pt depth0pt}
\def\IZ{\relax\ifmmode\mathchoice
{\hbox{\cmss Z\kern-.4em Z}}{\hbox{\cmss Z\kern-.4em Z}}
{\lower.9pt\hbox{\cmsss Z\kern-.4em Z}}
{\lower1.2pt\hbox{\cmsss Z\kern-.4em Z}}\else{\cmss Z\kern-.4em
Z}\fi}
\def\IB{\relax{\rm I\kern-.18em B}}
\def\IC{{\relax\hbox{$\inbar\kern-.3em{\rm C}$}}}
\def\ID{\relax{\rm I\kern-.18em D}}
\def\IE{\relax{\rm I\kern-.18em E}}
\def\IF{\relax{\rm I\kern-.18em F}}
\def\IG{\relax\hbox{$\inbar\kern-.3em{\rm G}$}}
\def\IGa{\relax\hbox{${\rm I}\kern-.18em\Gamma$}}
\def\IH{\relax{\rm I\kern-.18em H}}
\def\II{\relax{\rm I\kern-.18em I}}
\def\IK{\relax{\rm I\kern-.18em K}}
\def\IC{\relax{\rm I\kern-.18em C}}
\def\IR{\relax{\rm I\kern-.18em R}}

%%%%%%%%%%%%%%%%%%%%%%%%%%%%%%%%%%%%%%%%%%%%%%%%%%%%%%%%%%%%%%%%%%%%%%%%%%%
%Blackboard letters
%  The prehistoric version of this font is known as "msym". Many unfortunate
%  souls still have this old (and UGLY) ancestor of "msbm". Time to join the
%  modern world guys!

\font\blackboard=msbm10 \font\blackboards=msbm7
\font\blackboardss=msbm5
\newfam\black
\textfont\black=\blackboard
\scriptfont\black=\blackboards
\scriptscriptfont\black=\blackboardss
\def\blackb#1{{\fam\black\relax#1}}

%   Those truly poor slobs who have neither "msbm not "msym" fonts can
% substitute
%   the definition

%\def\blackb{\bf}

%   for the above font definitions or, if all else fails,
%   return to scratching symbols in the dirt with a sharpened stick.
%
\def\IC{\bf{\blackb C}}

%----------------------------
\lref\juan{J. Maldacena, ``The Large N Limit of Superconformal Field
Theories and Supergravity," hep-th/9711200.}
\lref\ads{S. Gubser, I. Klebanov and A. Polyakov, ``Gauge
Theory Correlators from Noncritical String Theory,'' hep-th/9802109\semi
E. Witten, ``Holography and Anti de Sitter Space,'' hep-th/9802150.}
\lref\ks{S. Kachru and E. Silverstein, ``4d Conformal Field Theories
and Strings on Orbifolds,'' Phys. Rev. Lett. {\bf 80} (1998) 4855,
hep-th/9802183.}
\lref\moore{G. Moore, ``Atkin-Lehner Symmetry,'' Nucl. Phys.
{\bf B293} (1987) 139.}
\lref\obs{See e.g. S. Weinberg, ``The Cosmological Constant Problem,''
Rev. Mod. Phys. {\bf 61} (1989) 1, and references therein.}
\lref\origsusy{Original SUSY papers}
\lref\ooguri{T. Eguchi and H. Ooguri, ``Chiral Bosonization on a Riemann
Surface,'' Phys. Lett. {\bf 187B} (1987) 127.}
\lref\tyeetal{G. Shiu and H. Tye,  ``Bose-Fermi Degeneracy and Duality
in Non-Supersymmetric Strings,'' hep-th/9808095.}
\lref\us{S. Kachru and E. Silverstein, ``Self-Dual Nonsupersymmetric
Type II String Compactifications,'' hep-th/9808056.}
\lref\origorb{L. Dixon, J. Harvey, C. Vafa, and E. Witten,
``Strings on Orbifolds,'' Nucl. Phys. {\bf B261} (1985) 678.}
\lref\vafadt{C. Vafa, ``Modular Invariance and Discrete Torsion
on Orbifolds,'' Nucl. Phys. {\bf B273} (1986) 592.}
\lref\freedvafa{D. Freed and C. Vafa, ``Global Anomalies on
Orbifolds,'' Comm. Math. Phys. {\bf 110} (1987) 349.}
\lref\fms{D. Friedan, E. Martinec and S. Shenker, ``Conformal Invariance,
Supersymmetry and String Theory," Nucl. Phys. {\bf B271} (1986) 93.}
\lref\dettheta{L. Alvarez-Gaume, G. Moore and C. Vafa, ``Theta Functions,
Modular Invariance and Strings,'' Comm. Math. Phys. {\bf 106} (1986) 1.}
\lref\amsgl{J. Atick, G. Moore and A. Sen, ``Some Global Issues
in String Perturbation Theory,'' Nucl. Phys. {\bf B308} (1988) 1.}
\lref\amscat{J. Atick, G. Moore and A. Sen, ``Catoptric Tadpoles,''
Nucl. Phys. {\bf B307} (1988) 221.}
\lref\Vpart{E. Verlinde and H. Verlinde, ``Multiloop Calculations
in Covariant Superstring Theory,'' Phys. Lett. {\bf B192} (1987) 95.}
\lref\Vcorr{E. Verlinde and H. Verlinde, ``Chiral Bosonization,
Determinants and the String Partition Function,'' Nucl. Phys.
{\bf B288} (1987) 357.}
\lref\dHp{E. d'Hoker and D. Phong, ``The Geometry of String Perturbation
Theory,'' Rev. Mod. Phys. {\bf 60} (1988) 917.}
\lref\lfact{O. Lechtenfeld, ``Factorization and Modular Invariance of
Multiloop Superstring Amplitudes in the Unitary Gauge,'' Nucl.
Phys. {\bf B338} (1990) 403.}
\lref\split{G. Falqui and C. Reina, ``A Note on the Global Structure
of Supermoduli Spaces,'' Comm. Math. Phys. {\bf 128} (1990) 247.}
\lref\edthree{E. Witten, ``Strong Coupling and the Cosmological
Constant,'' Mod. Phys. Lett. {\bf A10} (1995) 2153, hep-th/9506101.}
\lref\ds{M. Dine and E. Silverstein, ``New M-theory Backgrounds
with Frozen Moduli,'' hep-th/9712166.}
\lref\swspin{N. Seiberg and E. Witten, ``Spin Structures in
Superstring Theory,'' Nucl. Phys. {\bf B276} (1986) 272.}
\lref\mp{A. Morozov and A. Perelomov, ``Statistical Sums in
Superstring Theory: Genus 2,'' Phys. Lett. {\bf B199} (1987) 209.}
\lref\mor{A. Morozov, ``Two-Loop Statsum of Superstrings,''
Nucl. Phys. {\bf B303} (1988) 343.}
\lref\lp{O. Lechtenfeld and A. Parkes, ``On Covariant
Multi-loop Superstring Amplitudes,'' Nucl. Phys. {\bf B332} (1990) 39.}
\lref\lecht{O. Lechtenfeld, ``Factorization and Modular Invariance
of Multiloop Superstring Amplitudes in the Unitary Gauge,''
Nucl. Phys. {\bf B338} (1990) 403.}
\lref\mumford{D. Mumford, {\it Tata Lectures on Theta}, Vols. I,II
(Birkhauser, Basel, 1983).}
\lref\polchinski{J. Polchinski, ``Evaluation of the One Loop String
Path Integral,'' Comm. Math. Phys. {\bf 104} (1986) 37. }
\lref\mcclain{B. McClain and B. Roth, ``Modular Invariance for
Interacting Bosonic Strings at Finite Temperature,''
Comm. Math. Phys. {\bf 111} (1987) 539.}
\lref\martinec{E. Martinec, ``Nonrenormalization Theorems
and Fermionic String Finiteness,'' Phys. Lett. {\bf B171}
(1986) 189.}
\lref\mart{E. Martinec, ``Conformal Field Theory on a (Super)Riemann
Surface,'' Nucl. Phys. {\bf B281} (1987) 157.}
\lref\ars{J. Atick, J. Rabin and A. Sen, ``An Ambiguity in
Fermionic String Perturbation Theory,'' Nucl. Phys. {\bf B299}
(1988) 279.}
\lref\bkv{M. Bershadsky, Z. Kakushadze and C. Vafa,
``String Expansion as Large-N Expansion of Gauge Theories,''
hep-th/9803076\semi
M. Bershadsky and A. Johansen,
``Large N Limit of Orbifold Field Theories,''
hep-th/9803249.}
\lref\lnv{A. Lawrence, N. Nekrasov and C. Vafa,
``On Conformal Field Theories in Four-Dimensions,''
hep-th/9803015.}
\lref\tom{T. Banks, ``SUSY Breaking, Cosmology, Vacuum
Selection and the Cosmological Constant in String Theory,''
hep-th/9601151.}
\lref\holography{
G. 't Hooft, ``Dimensional Reduction in Quantum Gravity,''
gr-qc/9310026\semi
L. Susskind, ``The World as a Hologram,''
J. Math. Phys. {\bf 36} (1995) 6377, hep-th/9409089.}
\lref\usnext{Work in progress.}
\lref\jeff{J. Harvey, ``String Duality and Nonsupersymmetric Strings,''
hep-th/9807213.}
\lref\kallmor{R. Kallosh and A. Morozov, ``On Vanishing
of Multiloop Contributions to 0,1,2,3 Point
Functions in Green-Schwarz Formalism for
Heterotic String,'' Phys. Lett. {\bf B207} (1988) 164.}

\Title{\vbox{\baselineskip12pt\hbox{hep-th/9807076}
\hbox{LBNL-41932, SLAC-PUB-7875, SU-ITP-98/35, UCB-PTH-98/33}
}}
{\vbox{\centerline{
Vacuum Energy Cancellation in }\smallskip
\centerline{a Non-Supersymmetric String}}
}
\centerline{Shamit Kachru$^{1}$, Jason Kumar$^{2}$ and Eva Silverstein$^{3}$
}
\bigskip
\bigskip
\centerline{$^{1}$Department of Physics}
\centerline{University of California at Berkeley}
\centerline{Berkeley, CA 94720}
\smallskip
\centerline{and}
\smallskip
\centerline{Ernest Orlando Lawrence Berkeley National Laboratory}
\centerline{Mail Stop 50A-5101, Berkeley, CA 94720}
\medskip
\medskip
\centerline{$^{2}$ Department of Physics}
\centerline{Stanford University}
\centerline{Stanford, CA 94305}
\medskip
\medskip
\centerline{$^{3}$ Stanford Linear Accelerator Center}
\centerline{Stanford University}
\centerline{Stanford, CA 94309}
\bigskip
\medskip
\noindent
We present a nonsupersymmetric orbifold of type II string theory
and show that it has vanishing cosmological constant at the one
and two loop level.  We
argue heuristically that the cancellation may persist
at higher loops.

\Date{July 1998}
%\draftmode

\newsec{Introduction}

One of the most intriguing and puzzling pieces of data
is the (near-)vanishing of the cosmological constant
$\Lambda$ \obs.
Unbroken supersymmetry would ensure that perturbative quantum
corrections to the
vacuum energy vanish (in the absence of a U(1) D-term)
due to cancellations between bosonic and fermionic
degrees of freedom.  However,
although both bosons and fermions
appear in the low-energy spectrum, they are not related by
supersymmetry and this mechanism for cancelling $\Lambda$
is not realized.

Because string theory (M-theory) is a consistent
quantum theory which incorporates gravity, it is interesting
(and necessary) to see how string theory copes with the
cosmological constant.
In a perturbative
string framework, because the string coupling $g_{st}$
(the dilaton) is dynamical,
the quantum vacuum energy constitutes a potential for it.
So the issue of turning on a nontrivial string coupling is
related to the form of the vacuum energy in string theory.

In this paper we present a class of perturbative string models
in which supersymmetry is broken at the string scale but
perturbative quantum corrections to the cosmological constant
cancel.  We begin with a simple mechanism that ensures the
(trivial) vanishing of the 1-loop vacuum energy (as well as
certain tadpoles and mass renormalizations).  We then compute
the (spin-structure-dependent part of the) 2-loop partition function
and demonstrate that it vanishes.  This requires some analysis
of worldsheet gauge-fixing conditions, modular transformations,
and contributions from the boundaries of moduli space.
Examination of the general form of higher-loop amplitudes
suggests that they similarly may cancel and we next
present this argument. We are
unable to rigorously generalize our 2-loop calculation to
higher loops at this point
because of the complications of higher-genus moduli space.
We hope to be able to make the higher-genus result more
precise by using an operator formalism as will become clearer
in the text,
though we leave that for future work.

In addition we discuss how this model may fit into the
framework \ks\ relating conformal fixed lines/points in quantum
field theory to vanishing dilaton potentials/isolated
minima of the dilaton potential in string theory.
This provides hints as to where to look for more general
models with vanishing $\Lambda$.  In particular we will
be interested in models without
the tree level bose-fermi degeneracy that we have here,
as well as models in which the dilaton is stabilized.
We should note in this regard that instead of working in 4d perturbative
string theory as we do here, we could consider the same class of models
in 3d string theory and consider the limit of large $g_{st}$.
If the appropriate D-brane bound states exist in this theory
to provide Kaluza-Klein modes of an M-theoretic fourth dimension,
one could obtain in this way 4d M-theory vacua with vanishing
cosmological constant and no dilaton (in this way similar to
the scenario of \edthree, but here without the need for
3d supersymmetry).

We understand that a complementary set of models has been found
in the free fermionic description \tyeetal.  We would like
to thank Zurab
Kakushadze for pointing out
(and fixing) an error in our original
model as presented at Strings '98.

\newsec{Nonabelian Orbifolds and the 1-loop Cosmological Constant}

Consider the worldsheet path integral formulation of orbifold compactifications
\origorb.  In general one mods out by a discrete symmetry group
of the 10-dimensional string theory.  This group involves rotations
of the left and right-moving
worldsheet scalars $X_{L,R}^\mu$ and fermions $\psi_{L,R}^\mu$
as well as shifts of
the scalars $X_{L,R}^\mu$.  Here $\mu=1,\dots,10$ is a
spacetime $SO(9,1)$ vector index.  The worldsheet path integral at a given
loop order $h$ splits up into
a sum over different twist structures, in which the fields are twisted
by orbifold group elements in going around the various cycles of
the genus-$h$ Riemann surface $\Sigma_h$.
These twists must respect the homology relation

\eqn\homolconstraint{\prod_{i=1}^h a_ib_ia_i^{-1}b_i^{-1} = 1}
where $a_i$ and $b_i$ are the canonical 1-cycles on $\Sigma_h$.
In particular, at genus 1, one sums over pairs $(g,h)$ of
{\it commuting} orbifold space group elements $g$ and $h$.

\smallskip
\epsfbox{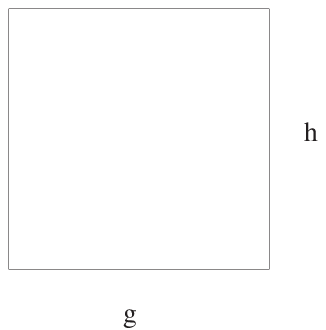}
\smallskip
\centerline{Figure 1: Torus twisted by elements $(g,h)$}
\smallskip

In considering nonsupersymmetric orbifolds, this suggests
an interesting class of models.  Consider orbifolds in which
no commuting pair of group elements breaks all the supersymmetry
(i.e. projects out all of the gravitinos), but in which
the full group does break all the supersymmetry.
At the one-loop level, each contribution to the path integral
then effectively preserves some supersymmetry and therefore
vanishes.  This is a formal way of encoding the fact that the spectrum
for this type of model will have bose-fermi degeneracy at
all mass levels (though no supersymmetry).  So the one-loop
partition function, as well as appropriate tadpoles,
mass renormalizations, and three-point functions, are uncorrected.

We will discuss the following specific model.\foot{Other similar
models can be constructed, some of which do not actually require
the group to be nonabelian to get 1-loop cancellation
\refs{\us,\tyeetal,\jeff}.}  Let us start with type II string
theory compactified on a square torus $T^6\sim (S^1)^6$ at the
self-dual radius $R=l_s$ (where $l_s=\sqrt{\alpha^\prime}$ is the
string length scale).  Consider the asymmetric orbifold generated
by the elements $f$ and $g$:
\bigskip
\vbox{\settabs 3 \columns
\+$S^1$&$f$&$g$\cr
\+1&$(-1,s)$&$(s,-1)$\cr
\+2&$(-1,s)$&$(s,-1)$\cr
\+3&$(-1,s)$&$(s,-1)$\cr
\+4&$(-1,s)$&$(s,-1)$\cr
\+5&$(s^2,0)$&$(s,s)$\cr
\+6&$(s,s)$&$(0,s^2)$\cr
\+&$(-1)^{F_R}$&$(-1)^{F_L}$\cr}

\noindent We have indicated here how each element acts on
the left and right moving RNS degrees of freedom of the
superstring. Here $s$ refers to a shift by $R/2$.  So for
example $f$ reflects the left-moving fields $X_L^{1\dots 4}$,
$\psi_L^{1\dots 4}$ and shifts $X_R^{1\dots 4}$ by $R/2$, $X_L^5$
by $R$, and $X^6={1\over 2}(X_L^6+X_R^6)$ by $R/2$.  In addition it
includes an action of $(-1)^{F_R}$ which acts with a $(-1)$ on all
spacetime spinors coming from right-moving worldsheet degrees of
freedom.  This can be thought of as discrete torsion \vafadt:  in
the right-moving Ramond sector the $f$-projection has the opposite
sign from what it would have without the $(-1)^{F_R}$ action.
Similarly the above table indicates the action of the generator $g$
on the worldsheet fields. This orbifold satisfies level-matching
and the necessary conditions derived in \refs{\vafadt,\freedvafa}
for higher-loop modular invariance (we do not know if these
conditions are sufficient).

There are several features to note about the spectrum of this model.
First, it is not supersymmetric.  In particular, $f$ projects out
all the gravitinos with spacetime spinor quantum numbers coming
from the right-movers.  Similarly $g$ projects out the gravitinos
with left-moving spacetime spinor quantum numbers.  Because
of the shifts included in our orbifold action, there are no
massless states in twisted sectors, so in particular no
supersymmetry returns in twisted sectors.  Second, the
model is nonetheless bose-fermi degenerate.  In particular the
massless spectrum has 32 bosonic and 32 fermionic physical
states.

In addition to the spectrum of perturbative string states
there is a D-brane spectrum in this theory which one
can analyze along the lines of \ds.This will be
of interest in placing this example in a more general context
in the final section.

Our orbifold group elements satisfy the following algebraic relations:
\eqn\orbalg{fg=gfT_L^{-1}T_R ~~~ fT_L^q=T_L^{-q}f ~~~ gT_R^q=T_R^{-q}g}
where $T_L$ denotes a shift by $R$ on $X_L^{1\dots 4}$ and
$T_R$ denotes a shift by $R$ on $X_R^{1\dots 4}$.  Clearly also $f$ commutes
with $T_R$ and $g$ commutes with $T_L$.

The first relation in \orbalg\ tells us that
$f$ and $g$ do not commute in the orbifold space group.  Therefore
at the one loop level they never both appear as twists $(f,g)$ in
the partition function (i.e. we cannot twist by $f$ on the
a-cycle and by $g$ on the b-cycle).  Furthermore we can check
that no commuting pair of elements break all the supersymmetry.
In order to break the supersymmetry we would need pairs of
the form $(fT_L^aT_R^b,gT_L^cT_R^d)$ or
$(fT_L^{\tilde a}T_R^{\tilde b}, fgT_L^{\tilde c}T_R^{\tilde d})$,
for arbitrary integers $a,b,c,d,\tilde a,\tilde b,\tilde c,
\tilde d$.
(We could also have the latter form with $f$ interchanged with $g$
but these are isomorphic.)
By using the relations \orbalg\ we see that neither pair of elements
commutes:
\eqn\noncommI{(fT_L^aT_R^b)(gT_L^cT_R^d)=
(gT_L^cT_R^d)(fT_L^aT_R^b)T_L^{2c+1}T_R^{1-2b}}
So there is no choice of integers $a,b,c,d$ for which the two elements
commute in the space group of the orbifold.  Similarly
\eqn\noncommII{
(fT_L^{\tilde a}T_R^{\tilde b})(fgT_L^{\tilde c}T_R^{\tilde d})
=(fgT_L^{\tilde c}T_R^{\tilde d})(fT_L^{\tilde a}T_R^{\tilde b})
T_L^{2\tilde c-2\tilde a-1}T_R^{1-2\tilde b}}
So at the one loop level, there will not be any contribution to
the partition function.

\newsec{The 2-loop vacuum energy}

At two loops the orbifold algebra itself does not automatically
ensure the cancellation of the partition function.  Let us
denote the canonical basis of 1-cycles by $2h$-dimensional
vectors $(a_1,\dots,a_h;b_1,\dots,b_h)$.  At genus two,
we run into twist structures like $(1,1;f,g)$ around the
canonical cycles:

\smallskip
\epsfbox{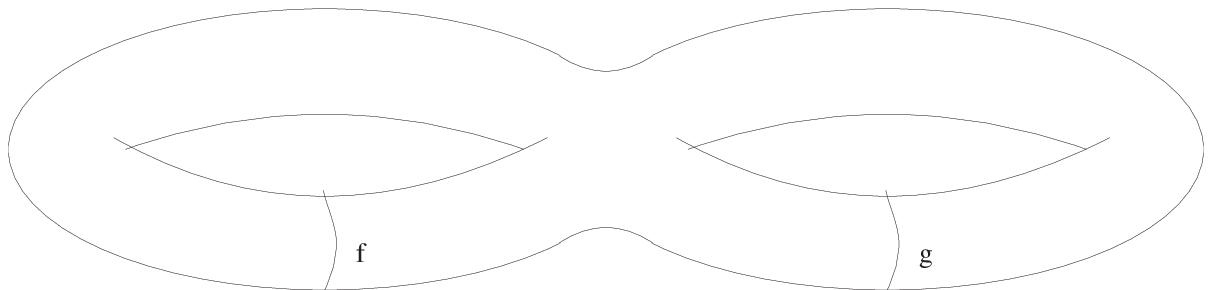}
\smallskip
\centerline{Figure 2: Basic twist structure at genus 2.}
\smallskip

In the figure we indicate the $\it cuts$ in the diagram in a given
twist structure--here the fields are twisted in going around
the $b$-cycles, as in doing so they pass through the indicated cuts.
In particular this diagram involves both $f$ and $g$ twists, and
therefore has the information about the full supersymmetry breaking
of the model.  Is there reason to believe the vacuum energy might
nonetheless cancel?  Heuristically, the following argument suggests
that we should indeed expect a cancellation.  Consider evaluating
the diagram of Figure 2 near the factorization limit in which
the diagram looks like a propagator tube connecting two tori.  Because
of the homology relations, in this twist structure the intermediate
state in this propagator is untwisted.  The diagram thus becomes
a sum over products of tadpoles of
untwisted propagating states (weighted by $e^{-mT}$ where
$m$ is the mass of the state and $T$ gives the length of
the tube).  Each term is
a tadpole of the untwisted state in the $g$-twisted theory
times a tadpole of the untwisted state in the $f$-twisted theory.
The contour deformation arguments of \fms\ imply that
these tadpoles vanish.  In order to make this rigorous one needs
to see explicitly that unphysical states decouple properly
(which only has to happen after summing over all twist
structures).
In what follows we will provide an explicit computation of
the 2-loop contribution and verify that it vanishes.

\subsec{Back to 1-loop.}

In order to appreciate the relevant mechanism, it is worth returning
momentarily to the 1-loop (supersymmetric) contribution $(1,f)$.
\smallskip
\epsfbox{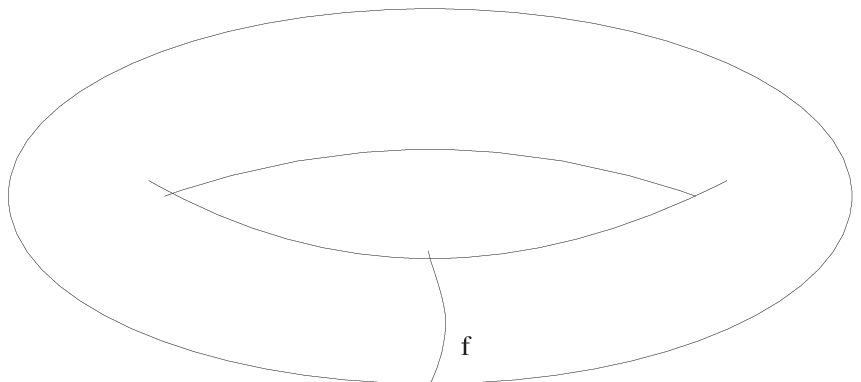}
\smallskip
\centerline{Figure 3: One-loop diagram with an $f$ twist on the $b$ cycle.}
\smallskip
This contribution must vanish by supersymmetry, but it is instructive
to observe how the spin structure sum works in this case before going
on to our 2-loop diagram.  The amplitude is
\eqn\ampI{{\cal A}_1=
\int {d^2\tau\over{(Im\tau)^2}}Tr(q^{L_0}\bar q^{\bar L_0}f)}
where $q=e^{2\pi i\tau}$ and $L_0$ and $\bar L_0$ are the
usual Virasoro zero mode generators.  Let us consider the spin-structure
dependent piece of this amplitude.  As explained in \dettheta, the
determinants for the worldsheet
Dirac operators acting on the RNS fermions are proportional to
theta functions.  The $\theta$-function is defined (for general genus $h$) by
\eqn\deftheta{\theta[\alpha,\beta](z|\tau )=
\sum_n e^{[\pi i(n+\alpha)^t\tau (n+\alpha)
+2\pi i(n+\alpha )(z+\beta )]}}
Here $z\in \IC^h/(\IZ^h+\tau \IZ^h)$ and $\tau$ is the period
matrix of the Riemann surface, defined in terms of the canonical
basis of holomorphic 1-forms $\omega_i$ by
$\oint_{a_j}\omega_i=\delta_{ij}$ and
$\oint_{b_j}\omega_i=\tau_{ij}$.
The characteristics $\alpha,\beta$ encode the spin structure \swspin, i.e.
the boundary conditions
of the fermions around the $a$ and $b$ cycles respectively of
the Riemann surface.  So for example if $\alpha_1=1/2$ (resp. 0), the
corresponding fermion has {\it periodic} (resp. {\it antiperiodic})
boundary conditions
around the $a_1$ cycle.

The integrand of the 1-loop amplitude \ampI\ is
proportional to
\eqn\thetaI{{\cal A}_1\propto\sum_{\alpha,\beta}
\eta_{\alpha,\beta}
\theta^2[\alpha,\beta](0|\tau)\theta^2[\alpha,\beta+{1\over 2}](0|\tau)}
where $\eta_{\alpha,\beta}$ are the phases encoding the GSO
projection.  The first $\theta^2$ factor comes from the left-moving RNS
fermions $\psi_L^{1\dots 4}$ and the second $\theta^2$ factor comes
from the other four transverse left-moving fermions
$\psi_L^{5\dots 8}$.  The symmetry between these two factors
will play an important role for us.  Let us consider first
the terms in the sum \thetaI\ with $\alpha=1/2$.  This describes
left-moving Ramond-sector states propagating in the loop, as the left-moving
fermions $\psi_L$
are periodic around the $a$-cycle.  Because we have
an $f$-twist around the $b$-cycle, half the $\psi_L^\mu$ are periodic
around the $b$-cycle and half are antiperiodic around the $b$-cycle
for each value of $\beta$ in the sum.  Thus in each $\alpha=1/2$
term half
the RNS fermions have zero modes, so these terms identically vanish.

Let us now consider the terms with $\alpha=0$, which describe
left-moving Neveu-Schwarz states propagating in the loop.  These give
\eqn\lastsum{\sum_{\beta=0,1/2}\eta_{0,\beta}
\theta^2[0,\beta](0|\tau)\theta^2[0,\beta+1/2](0|\tau).}
Note that both terms in this sum have the same functional
form ($\theta^2[0,1/2](0|\tau)\theta^2[0,0](0|\tau)$).  The only issue
left is then the relative phase between them.  The
sum over $\beta$ is simply the GSO projection on
the states propagating around the $b$-cycle. Let us
normalize $\eta_{0,0}$ to 1.  Then $\eta_{0,1/2}=-1$.
This follows from the fact that {\it in the NS sector} the
GSO projection operator is $1-(-1)^F$.  This encodes the
fact that we must project onto odd fermion number in the
superstring in order to project out the tachyon which
would otherwise come from the vacuum at the $-1/2$ mass level.
So our integrand is
\eqn\finI{(1-1)\theta^2[0,0]\theta^2[0,1/2]=0.}

\subsec{2-loops with supersymmetry}

In order to proceed to the 2-loop computation, we must consider various
subtleties arising in string loop computations for strings
with worldsheet supersymmetry.  (See for example \refs{\amsgl,\amscat,
\dHp}
for reviews with some references.)  Let us begin by briefly
reviewing some of the issues in the supersymmetric case.  We will work
in the RNS formulation;
for discussion of the supersymmetric case in Green-Schwarz language
see for example \kallmor.

In performing the Polyakov path integral at genus $h$, we
must integrate over all the worldsheet fields including
the worldsheet metric $\hat h$ and gravitino
$\chi$.  This infinite dimensional space is reduced to
a finite dimensional space of (super-)moduli by dividing
out the diffeomorphisms and local supersymmetry transformations.
There are $3h-3$ complex bosonic moduli $\tau$ and $2h-2$ complex
supermoduli $\zeta$.  At genus $h=2$ we can take the gravitino
to have delta-function support on the worldsheet for even
spin structures \split.  (For odd spin structures the amplitude
vanishes as a result of the integration of fermionic zero modes.)

We will review the supersymmetric cancellation at
2 loops.  As explained for example in \refs{\dHp,\Vpart,\amsgl},
the type II string path integral can be written as
\eqn\path{
\int_{{\cal SM}_h}d\mu_0\int [dBdCdX]e^{-S}
\prod_{r=1}^{6h-6}(\eta_r,B)
\prod_{a=1}^{4h-4}\delta((\eta_a,B))
} Here $B,C$ denote the $b,\beta$ and $c,\gamma$ ghosts, where
$(b,c)$ are the spin-(2,-1) conformal ghosts and $(\beta,\gamma)$
are the spin-(3/2,-1/2) superconformal ghosts.  $X$ denotes the
matter fields and $\eta_r$ and $\eta_a$ are Beltrami differentials
relating the metric and gravitino to the moduli and supermoduli (in
essence, they determine the way in which superdiffeomorphism
invariance is gauge-fixed). In components,
\eqn\beltI{
(\eta_r,B)=\int\eta_{r\bar z}^z b_{zz}+\int\eta_{r\bar z}^+ \beta_{z+}
+\int\eta_{r z}^{\bar z} b_{\bar z\bar z}
+\int\eta_{r z}^- \beta_{\bar z -}
}
\eqn\beltII{
(\eta_a,B)=\int\eta_{a\bar z}^z b_{zz}+\int\eta_{a\bar z}^+ \beta_{z+}
+\int\eta_{a z}^{\bar z} b_{\bar z\bar z}
+\int\eta_{a z}^- \beta_{\bar z -}
}
As explained e.g. in \refs{\mart,\dHp},
we can write the path integral measure on
supermoduli space in terms of
a fixed measure on
moduli space
\eqn\detch{
d\mu_0=d\mu [sdet(\eta,\Phi)]^{-1}[sdet(\Phi,\Phi)]^{1/2}
}
Here $d\mu$ is a fixed measure on the supermoduli space
${\cal{SM}}_h$, integrated over a fixed domain independent
of the beltrami differentials.  $\Phi$
contains the $3h-3$ holomorphic and $3h-3$ antiholomorphic
2-differentials ($b$ ghost zero mode wavefunctions) and
the $2h-2$ holomorphic and $2h-2$ antiholomorphic 3/2-differentials
($\beta$ ghost zero modes).

After choosing delta-function support for the worldsheet gravitinos,
and integrating out the supermoduli, one obtains a correlation
function of picture changing operators \Vpart
\eqn\picop{
:e^\phi T_F: = c\partial\xi + {1\over 2} e^\phi\psi^\mu\partial X^\mu
-{1\over 4}\partial\eta e^{2\phi}b
-{1\over 4}\partial (\eta e^{2\phi} b)}
and other ghost insertions
\eqn\part{\eqalign{
\sum_{\alpha,\beta,twists}\int d\mu
[sdet(\eta,\Phi)]^{-1}[sdet(\Phi,\Phi)]^{1/2}[dX][dB][dC]e^{-S}
(\hat\eta,b)^{6h-6}\xi (x_0)\cr
\prod_{a=1}^{2h-2}:e^\phi T_F(z_a):
\prod_{a=2h-3}^{4h-4}:e^{\bar\phi}\bar T_F(z_a):\bar\xi(\bar y_0)\cr}}
The superconformal ghosts $\beta=\partial\xi e^{-\phi}$,
$\gamma=\eta e^{\phi}$ are defined in terms of spin-0 and
spin -1 fermions $\xi,\eta$ and a scalar $\phi$ \fms.
The spin-0 fermion $\xi$ has a zero mode on the surface which
is absorbed by the insertion of $\xi (x_0)$ in \part.
There is an anomaly in the ghost number $U(1)$ current which
requires insertions of operators with total ghost number
$2h-2$ to get a nonvanishing result.
The correlation functions \part\ can be evaluated using the formulas
derived in e.g. \refs{\Vcorr,\ooguri}.

%.
%On our genus $g=2$ Riemann surface $\Sigma$, we have $2^g$ cycles $a_i$
%and $b_i$.  It can be shown that the basis of holomorphic 1-forms on
%$\Sigma$ can be given by
%
%\eqn\holoforma{
%\oint_{a_i}^{}{\omega_j}= \delta_{ij}
%}
%\eqn\holoformb{
%\oint_{b_i}^{}{\omega_j}= \tau_{ij}
%}
%where $\tau_{ij}$ is the period matrix of the Riemann surface.  We
%choose a base point $p_0$ on $\Sigma$, and define the Jacobi map
%which takes a divisor on the Riemann surface (in this case, essentially
%a formal sum of points) to
%
%\eqn\Jacobi{I(\sum {p_i} - \sum{q_j}) = \sum {\int_{p_0}^{p_i}{\omega_i}}
%-\sum {\int_{p_0}^{q_j}{\omega_i}}}
%defined on the Jacobi torus $J(\Sigma)$.
%Since the domain of theta functions is the Jacobi torus,
%$\theta[\delta](z_1 - z_2)$ is actually a shorthand for
%$\theta[\delta](\int_{p_0}^{z_1}{\omega} - \int_{p_0}^{z_2}
%{\omega})$.  The Riemann Vanishing Theorem states for a spin
%structure $\delta$ and a base point $p_0$, there exists a degree
%$g-1$ divisor class $\Delta_\delta$, such that
%$\theta[\delta] (z)=0$ if and only if there exist points
%$p_1$ ,...,$p_{g-1} \in \Sigma$ such that
%
%\eqn\RVT{I(z)=I(\Delta_{\delta})-I(\sum_{i=1}^{g-1}{p_i})}
%
%As a particular example of the usefulness of the Riemann Vanishing
%Theorem, suppose that $\delta$ is an odd spin sturcture (for simplicity,
%we now specify $g=2$).  Since $I(\Delta_{\delta}
%-z)=I(\Delta_\delta) - I(z)$, we conclude that $\theta[
%\delta](z-\Delta_\delta) = -\theta[\delta](\Delta_\delta - z)=0$ for all
%$z\in \Sigma$, by
%the Riemann Vanishng Theorem.
%
We will  now fix the gauge for the gravitinos by making a definite
choice of points $z_{1,2}$.  As explained in \amsgl,
the choice of points must be taken in such a way that
the gauge slice chosen is transverse to the gauge transformations.
It must also respect modular invariance of the amplitude
\refs{\Vpart,\amsgl}.
Ultimately, we will be interested in a gauge choice for which
$z_1,z_2\to\Delta_\gamma$, where
$\Delta_\gamma$ is a divisor corresponding to an
odd spin structure $\gamma$, that is a point where a holomorphic
1/2-differential has a zero.
As explained in \amsgl, this choice
(which amounts to putting the insertions at one of the branch points
in a hyperelliptic description of the surface) satisfies transversality.
It was argued in \refs{\lp,\lecht}
that despite earlier worries \amsgl, this
choice is also consistent with modular invariance.  The modular
invariance is not manifest in the description in terms
of $\theta$-functions, as the calculation of correlation
functions on the Riemann surface \Vcorr\ involve a choice
of reference spin structure $\delta$.  Having to choose a spin
structure naively appears to violate modular invariance.
Had we chosen a different reference spin structure $\delta^\prime$,
we would have shifted the arguments of our theta functions
by elements $n+m\tau$ of the Jacobian lattice.
Such a shift introduces a $\tau$-dependent phase multiplying
the $\theta$-function--the $\theta$-functions transform as
sections of line bundles over the Jacobian torus.  These
phases must cancel out of the properly defined integrand,
and in \lp\ this was demonstrated explicitly for certain (nonvanishing)
2-loop contributions.

We need to consider the (left-moving) spin-structure-dependent pieces
of the correlation function, the poles arising from
the spin-structure-independent local behavior of
the picture changing correlator, and the behavior of the
determinant \detch\ in this gauge.
According to \Vpart\ we have the following contributions
to the spin-structure-dependent pieces of the 2-loop partition function.
The matter part of $T_F$ contributes

\eqn\matt{
\sum_\delta \langle \delta|\gamma \rangle
{{\theta[\delta]^4(0)\theta[\delta](z_1-z_2)}\over
{\theta[\delta](z_1+z_2-2\Delta_\gamma)}}}
Here $\delta\equiv (\alpha,\beta )$ encodes the
spin structure of the various contributions and
$\langle \delta|\gamma \rangle = e^{4\pi i(\alpha\gamma_2-\beta\gamma_1)}$
encodes the GSO phases \lfact.  Here the arguments $\sum p-\sum q$
in terms of $p$ and $q$ which are sets of points on the Rieman surface is
shorthand for the Jacobi vector $\sum\int_{p_0}^p\omega_i
-\sum\int_{p_0}^q\omega_i$.

Let us first, following \lp, take $z_1+z_2=2\Delta_\gamma$,
that is place $z_1+z_2$ at a divisor corresponding to the
canonical class,
without setting $z_1=z_2$.
The contribution \matt\ then simplifies to
\eqn\nummatt{
\sum_\delta \langle \delta|\gamma \rangle \theta[\delta]^3(0)\theta[\delta](z_1-z_2)
=4\theta[\gamma]^4(z_1-\Delta_\gamma)
}
where in the last step we have used a Riemann identity.
The Riemann Vanishing Theorem then implies that
this vanishes identically as a function of $z_1$ \lp.
Thus in this case whatever poles arise as $z_1\to z_2$,
the identical zero from the spin structure sum cancels it.

Now turning to the ghost piece of the correlation function of
picture-changing operators, one obtains contributions
isomorphic to \nummatt\ as well as
\eqn\ghost{
\omega_i(z_1){{\theta[\delta]^5(0)\partial_i\theta[\delta](2z_2-2\Delta_\gamma)}
\over{\theta^2[\delta](z_1+z_2-2\Delta_\gamma)}}
}
Here $\omega_i$ are the canonical basis of holomorphic
one-forms on the Riemann surface, satisfying
$\int_{a_i}\omega_j=\delta_{ij}$ and
$\int_{b_i}\omega_j=\tau_{ij}$ where $\tau$ is the
period matrix for the surface.
Again simplifying this by first taking
$z_1+z_2=2\Delta_\gamma$ we obtain
\eqn\numghost{
\sum_\delta \langle \delta|\gamma \rangle
\partial_{z_1}(\theta[\delta]^3(0)\theta[\delta](z_1-z_2))
=4\partial_{z_1}(\theta[\gamma]^4(z_1-\Delta_\gamma))}
Because the right-hand side of this expression is a
derivative of 0 (by the Riemann vanishing theorem), it
vanishes identically.  Again any poles from the picture changing
OPEs are irrelevant \lp.

\subsec{(Non-)Superstring Perturbation Theory}

In an orbifold model, one can consider separately different
twist structures, and analyze the fundamental domain of
the modular group that preserves a given twist structure.
In general there are an infinite number of contributions coming
from different choices of bosonic shifts.  In $\S4$
we will analyze the twist structure of Figure 2 (with no
additional bosonic shifts) and see that the
resulting modular group acts freely on $\tau$.
In this situation, the choice of a branch point for $z_{1,2}$
is manifestly modular invariant; the possible obstruction to modular invariance
discussed in \refs{\Vpart,\amsgl} does not arise, as there are no
orbifold points in the moduli space.
We also analyze in $\S4$ the boundary contributions and see that they
vanish.
One can show that with arbitrary additional shifts (respecting the
homology relation of the Riemann surface) there are still no
orbifold points in the moduli space.

We will analyze the twist structure
$(1,1,f,g)$ (it will later be shown why this is the only twist
structure which needs to be analyzed).
The $f$ twist affects the characteristics of
some of the $\theta$ functions (arising from twisted fields)
by shifting them by $(0,0,1/2,0)$ -- we shall denote this
as a shift by ${1\over 2}L$.  $\kappa$ will be defined as
$\gamma + (0,0,0,1/2)$, and we choose $\gamma$ such
that both $\gamma$ and $\kappa$
are odd.

The correlation function of
the matter part of the picture changing operators breaks into
two contributions.  The terms involving
$\langle \psi^i\partial X^i(z_1)\psi^i\partial X^i(z_2) \rangle$ with
$i=5,\cdots 10$ give
\eqn\mattl{\eqalign{
\sum_\delta \langle \kappa|\delta \rangle
{{\theta[\delta](0)^2\theta[\delta+{1\over 2}L](0)^2\theta[\delta](z_1-z_2)}
\over{\theta[\delta](z_1+z_2-2\Delta_\gamma)}}\cr
\times ( p_i^\mu\omega^i(z_1)p_j^\mu\omega^j(z_2)
{1\over{E(z_1,z_2)^2}}
+{6\over{E(z_1,z_2)^2}}\partial_{z_1}\partial_{z_2}log E(z_1,z_2))\cr
\times det(\Phi^{3/2}_a(z_b))}}

Upon setting $z_1+z_2=2\Delta_\gamma$ we can cancel the denominator
against one factor in the numerator to get
\eqn\mattsimp{
\sum_\delta \langle \kappa |\delta \rangle
\theta[\delta](0)\theta[\delta](z_1-z_2)
\theta[\delta+{1\over 2}L](0)^2
=4\theta[\kappa]({1\over 2}(z_1-z_2))
\theta[\kappa+{1\over 2}L]({1\over 2}(z_1-z_2))
}
for the spin-dependent piece of this correlator.
Because $\kappa$ is an odd spin structure, this vanishes like
$(z_1 - z_2)^{2}$.  As $z_1\to z_2$ the determinant
factor \detch\ produces another zero:
plugging in the delta function $\eta_a$ we obtain
\eqn\detzer{
[sdet(\eta,\Phi)]^{-1}\propto
det(\eta_a,\Phi_b^{3/2})
=det(\Phi_b^{3/2}(z_a))}
Here $\Phi_b^{3/2}, b=1,2$ form a basis of holomorphic 3/2-differentials.
As the $z_a$ approach each other, the determinant \detzer\ goes
to zero, so
all in all \mattl\ has a $(z_1-z_2)^3$ multiplying the
prime forms.  However, since the prime-forms are yielding poles
as $z_1 \to z_2$, it remains to check that there are no finite
pieces in \mattl.

Note that $E(z_1,z_2)$ goes like $z_1-z_2$ as $z_1\to z_2$.
Thus, the terms proportional to ${1\over E(z_1,z_2)^2}$ times the loop
momenta clearly vanish in
the limit, since there is only a second order pole from the prime forms
which cannot cancel the third order zero we found from the spin structure
sum and the superdeterminant.  This leaves the term which goes like
${1\over E(z_1,z_2)^2} \partial_{z_1}\partial_{z_2}log E(z_1,z_2)$.
Using the fact that $E(z_1,z_2)$ has a Taylor expansion of the form
\eqn\eexp{E(z_1,z_2) \sim \sum_{n=0}^{\infty} c_n (z_1 - z_2)^{2n+1}}
as $z_1 \to z_2$, one sees that this combination of prime forms has
an expansion
\eqn\pexp{{1\over E(z_1,z_2)^2} \partial_{z_1} \partial_{z_2} log E(z_1,z_2)
\sim {\sum_{n=-2}^{\infty} d_{n} (z_1-z_2)^{2n}}}
On the other hand, the determinant factor is an $\it odd$ function of
$z_1 - z_2$ with an expansion of the form
\eqn\detexp{det(\Phi^{3/2}_a(z_b)) \sim \sum_{m=0}^{\infty} e_m (z_1-z_2)^{2m+1}}
while the sum over spin structures \mattsimp\ is an even function with a
second order zero at $z_1=z_2$.  From these facts, it is easy to see that
the full expression \mattl\ has an expansion of the form
\eqn\mattexp{\sum_{j=0}^{\infty} f_{j} (z_1-z_2)^{2j-1}}
as $z_1 \to z_2$.

Examining \mattexp, we see that

\noindent
$\bullet$ There are no finite contributions as $z_1 \to z_2$.

\noindent
$\bullet$ There is a (gauge artifact) pole as $z_1 \to z_2$; in fact
this pole receives contributions from the various matter and ghost correlators
proportional to the matter/ghost central charges, and hence cancels once
all of the terms are taken into account (since $c_{tot} =
c_{matter} + c_{ghost}=0$).  We will see this explicitly once we
compute the remaining matter and ghost contributions.

The second type of matter correlator arises from contracting
the $\psi^{i}\partial X^{i}(z_1) \psi^{i}\partial X^i (z_2)$ with
$i=1,\cdots 4$.
This leads to a contribution
\eqn\mattwo{\eqalign{\sum_{\delta}\langle \kappa | \delta \rangle {{\theta[\delta](0)^3
\theta[\delta + {1\over 2}L](0) \theta[\delta + {1\over 2}L] (z_1-z_2)}\over
{\theta[\delta](z_1 + z_2 - 2\Delta_{\gamma})}}\cr
\times (p_i^\mu \omega^i (z_1) p_j^\mu \omega^j (z_2) {1\over E(z_1,z_2)^2} +
{4\over E(z_1,z_2)^2}\partial_{z_1}\partial_{z_2} log E(z_1,z_2))\cr
\times det(\Phi^{3/2}_a(z_b))}}

Choosing $z_1 + z_2 = 2\Delta_{\gamma}$, the spin sum in \mattwo\
simplifies to
\eqn\spintwo{\sum_{\delta} \langle \kappa | \delta\rangle \theta[\delta](0)^2
\theta[\delta + {1\over 2}L](0) \theta[\delta + {1\over 2}L] (z_1-z_2)}
which, after applying a Riemann identity, becomes
\eqn\spintwon{4 \theta[\kappa]({1\over 2}(z_1-z_2))^{2}
\theta[\kappa+{1\over 2}L]
({1\over 2} (z_1-z_2))^2}
So in fact after summing over spin structures this looks the same as the
spin sum of the first type of matter contribution \mattsimp.
Again, it vanishes like $(z_1-z_2)^2$ as $z_1 \to z_2$.

Now, the argument for the cancellation proceeds as it did for the first
type of matter contribution.
The terms involving only the ${1\over E(z_1,z_2)^2}$ multiplying loop
momenta only have a second order pole, which cannot cancel the third
order zero coming from the determinant times the spin structure sum \spintwon.
The terms involving higher inverse powers of the prime forms lead to
a simple pole (which cancels after summing over matter and ghosts, as it is
proportional to the total central charge) and no finite contributions.

Next, let us consider the terms in the correlator of picture changing
operators coming from the ghost part of the worldsheet supercurrent
These terms take the form
\eqn\ghostcorr{\eqalign{
\langle -{1\over 4} c\partial\xi(z_1) \left(2 \partial \eta e^{2\phi}b
+ \eta \partial e^{2\phi} b + \eta e^{2\phi}\partial b \right)(z_2)\rangle \cr
+
\langle -{1\over 4} \left(2 \partial \eta e^{2\phi} b +
\eta \partial e^{2\phi} b + \eta e^{2\phi} \partial b \right)(z_1) c\partial
\xi (z_2)\rangle~.\cr}}

There are three types of terms that arise \Vpart.
We are in the twist structure $(1,1,f,g)$.  As in the matter
sector, the $f$ twist affects the characteristics of the $\theta$-functions
arising in the worldsheet correlation functions and determinants.
We will denote the shift in the characteristic, which is
$(0,0,1/2,0)$, as ${1\over 2}L$.
The
first type of contribution is
\eqn\firstgh{\eqalign{
\sum_\delta \langle \kappa |\delta \rangle
{{\theta[\delta](0)^3\theta[\delta+{1\over 2}L](0)^2
\theta[\delta](2z_2-2\Delta_\gamma)\theta(z_1-z_2+\sum w-3\Delta)}
\over{\theta[\delta](z_1+z_2-2\Delta_\gamma)^2 E(z_1,z_2)^3}}\cr
\times det (\Phi_a(z_b)) {{\prod E(z_1,w)}\over{\prod E(z_2,w)}}
\partial_{z_1}log({{\prod E(z_1,w)}\over{E(z_1,z_2)^5\sigma(z_1)}})
\cr
+(z_1\leftrightarrow z_2)\cr}}
The second is
\eqn\secondgh{
\eqalign{
\sum_\delta \langle \kappa |\delta \rangle
{{\theta[\delta](0)^3\theta[\delta+{1\over 2}L](0)^2
\omega_i(z_1)\partial^i\theta[\delta](2z_2-2\Delta_\gamma)
\theta(z_1-z_2+\sum w-3\Delta)}
\over{\theta[\delta](z_1+z_2-2\Delta_\gamma)^2 E(z_1,z_2)^3}}\cr
\times det (\Phi_a(z_b)) {{\prod E(z_1,w)}\over{\prod E(z_2,w)}}
\cr
+(z_1\leftrightarrow z_2)\cr
}}
The third is
\eqn\thirdgh{
\eqalign{
\sum_\delta \langle \kappa |\delta \rangle
{{\theta[\delta](0)^3\theta[\delta+{1\over 2}L](0)^2
\theta[\delta](2z_2-2\Delta_\gamma)
\omega_i(z_1)\partial^i\theta(z_1-z_2+\sum w-3\Delta)}
\over{\theta[\delta](z_1+z_2-2\Delta_\gamma)^2 E(z_1,z_2)^3}}\cr
\times det \Phi_a(z_b) {{\prod E(z_1,w)}\over{\prod E(z_2,w)}}
\cr
+(z_1\leftrightarrow z_2)\cr
}}

Setting $z_1+z_2=2\Delta_\gamma$ and doing the spin structure
sum we find for the spin-structure-dependent pieces of
contributions \firstgh\ and \thirdgh :
\eqn\spsumI{\eqalign{
\sum_\delta \langle \kappa |\delta \rangle
\theta[\delta](0)\theta[\delta+{1\over 2}L](0)^2
\theta[\delta](2z_2-2\Delta_\gamma)\cr
=\theta[\kappa](z_2-\Delta_\gamma)^2
\theta[\kappa+{1\over 2}L](z_2-\Delta_\gamma)^2\cr
\sim (z_1-z_2)^2+c_4(z_1-z_2)^4+\dots\cr}}
for some constant $c_4$
where in the last line we expanded the result in
a Taylor expansion around $z_1=z_2$.
For contribution \secondgh\ we get
\eqn\spsumII{\eqalign{
\sum_\delta \langle \kappa |\delta \rangle
\theta[\delta](0)\theta[\delta+{1\over 2}L](0)^2
\omega_i(z_1)\partial^i\theta[\delta](2z_2-2\Delta_\gamma)\cr
=
\partial_{z_1}\biggl( \theta[\kappa]({1\over 2}(z_1-z_2))^2
\theta[\kappa+{1\over 2}L]({1\over 2}(z_1-z_2))^2\biggr)\cr
\sim (z_1-z_2)+b_3(z_1-z_2)^3+\dots\cr }}

As for the matter contributions, although the spin structure
sums give vanishing contributions, they multiply singularities
arising from the prime forms $E(z_1,z_2)$ and we must analyze
the potential finite terms in the Taylor expansion.
Let us consider first \firstgh.  There are two types of
contributions here.  After doing the spin structure
sum as above the first takes the form
\eqn\firstform{\eqalign{
-5{{\partial_1 E(z_1,z_2)}\over{E(z_1,z_2)^4}}[z_{12}^2+c_4z_{12}^4+\dots]
[z_{12}+e_3z_{12}^3+\dots]{{\prod E(z_1,w)}\over{\prod E(z_2,w)}}
+(z_1\leftrightarrow z_2)\cr}}
where we denote $z_1 - z_2$ by $z_{12}$.
Here the second factor comes from the spin structure sum, the
third from the Taylor expansion of the determinant about
$z_1=z_2$ (where $e_3$ is some constant).
We should emphasize what is meant here by $(z_1\leftrightarrow z_2)$.
We are computing a correlation function of picture changing
operators.  The ghost piece of this correlator has the form
\ghostcorr.  So for example the second term in \ghostcorr\ corresponds to the
term denoted $z_1\leftrightarrow z_2$ in \firstform.  So
in particular the second term involves interchanging the operators
in the ghost correlator, without changing $z_1$ to $z_2$ in the
determinant factor.
The first and fourth
factors involving the prime forms encode the physical poles
and zeroes of the correlator.
The leading singularity from the prime forms here comes from
the $1/z_{12}^4$ term in the expansion of the prime form factors.
Therefore only the leading term in the Taylor expansion of the
spin structure sum and determinant factors potentially survive
(so we can ignore the terms proportional to $c_4$ or $e_3$,
which give fifth-order zeroes).
Similarly expanding the prime forms $E(z_1,z_2)$ gives a subleading
term with only a $1/z_{12}^2$ pole, which is cancelled by the third
order zero coming from the leading piece of the spin structure
sum times determinant.

Putting the factors together, we see that the leading piece is
a simple pole in $z_{12}$.
The first three factors in \firstform\
are the same in the term with $z_1\leftrightarrow z_2$.
When we include the term with $z_1\leftrightarrow z_2$, they
multiply the
prime form factor
${{\prod E(z_1,w)}\over{\prod E(z_2,w)}}+
{{\prod E(z_2,w)}\over{\prod E(z_1,w)}}$.  This is
even under $z_1\leftrightarrow z_2$.  In our Taylor expansion
it therefore becomes of the form ${\cal O}(1)+f_2z_{12}^2+\dots$,
and only the first term contributes.
Therefore in Taylor expanding the contribution \firstform, we
get a pole piece
plus higher order terms which vanish in the
limit $z_1\to z_2$.  In particular, no finite pieces survive.
What is the interpretation of the pole piece?  It is proportional to
the ghost central charge, and precisely cancels the pole piece
coming from the matter contribution.

The second type of contribution in \firstgh\ takes the form
\eqn\firstformb{\eqalign{
{1\over{E(z_1,z_2)^3}}[z_{12}^2+\dots][z_{12}+\dots]
{{\prod E(z_1,w)}\over{\prod E(z_2,w)}}\partial_1 log({{\prod E(z_1,w)}
\over{\sigma(z_1)}})
\cr}}
where the $\dots$ denotes terms which vanish automatically as
$z_1\to z_2$.  The leading pole from the prime forms here is
cubic.  Before including the $z_1\leftrightarrow z_2$ term there
is a finite piece obtained by multiplying this times the third
order zero obtained from the spin structure sum and determinant
factors.  The spin structure sum is even under the interchange
of $z_1$ and $z_2$ in this case, and as discussed above the
determinant factor is the same in both terms.  The factor
${1\over{E(z_1,z_2)^3}}$ does change sign between the two
terms, however.  So when we add the $(z_1\leftrightarrow z_2)$ term
the contribution cancels.

Let us now consider the contribution \secondgh.  This gives a
contribution of the form
\eqn\secondform{
{1\over{E(z_1,z_2)^3}}[z_{12}+\dots][z_{12}+\dots]
\biggl({{\prod E(z_1,w)}\over{\prod E(z_2,w)}}+
{{\prod E(z_2,w)}\over{\prod E(z_1,w)}}\biggr)
}
Here similarly to the above analysis we took into account the
relative sign of the two contributions in \ghostcorr\ and included
the $z_1\leftrightarrow z_2$ contribution.  The last factor
here is even under interchange of $z_1$ and $z_2$, so its
Taylor expansion is of the form $1+h_2z_{12}^2+\dots$ for
some constant $h_2$.  The leading contribution here is
a simple pole, and
there is no finite contribution.

Unlike the previous simple poles we have encountered, the pole
encountered here does not cancel with the other matter and ghost
contributions (it is $\it not$ one of the pieces which would have
contributed to the ${c\over z^{4}}$ pole in the OPE of
picture changing operators before accounting for
spin structure sums and determinant factors).
However, on general grounds we expect such gauge artifact poles to
constitute total derivatives on moduli space.
Otherwise the invariance of the path integral on gauge slice
would be lost.  In this case, we can
argue for that conclusion as follows.  The pole we are discussing
receives a ${1\over {(z_1-z_2)^{3}}}$ contribution from the prime forms
which is softened to ${1\over {(z_1-z_2)^{2}}}$ by the theta function
zero (and then to a simple pole by the determinant factor).
In the OPE of picture changing operators, the
${1\over {(z_1-z_2)^2}}$ divergence is multiplied by the stress-energy
tensor, which gives a derivative with respect to the metric and therefore
the moduli.  The term we
are finding is part of this total derivative.  In the gauge we have
chosen, it is the only non-vanishing piece (the other pieces vanish
even before integration over the moduli space).  However, since there
cannot be gauge artifact poles, we expect it to integrate to zero (which one
can argue for by analyzing the boundary contributions, as we will do later).

Finally let us consider the last ghost contribution \thirdgh.
This contribution takes the form
\eqn\thirdform{\eqalign{
{1\over{E(z_1,z_2)^3}}[z_{12}^2+\dots][z_{12}+\dots]
{{\prod E(z_1,w)}\over{\prod E(z_2,w)}}\cr
+(z_1\leftrightarrow z_2)\cr
}}
In this contribution before including the
$z_1\leftrightarrow z_2$ contribution there is a
potential finite term from the third order pole
multiplying a third order zero in $z_{12}$.
Here again, in the limit $z_1\to z_2$
every factor except the first is the same in
the two terms.  The first factor ${1\over{E(z_1,z_2)^3}}$ has
the opposite sign in the two terms.  Thus again after including
the $z_1\leftrightarrow z_2$ term the contribution cancels.

\newsec{Boundary Contributions}

In the previous section, we studied the two loop diagram with
twists by $f$ and $g$ going around the $b_{1,2}$ cycles, i.e.
with twist structure $(1,1,f,g)$.  We saw that the
computation yields a $\it vanishing ~integrand$ if we make a
very specific choice of insertion points for the picture-changing
operators $:e^{\phi}T_F:$.  Since the answer should be independent
of the choice of these insertion points, this seems to imply that
the two loop vacuum energy vanishes.

However, under a change of the choice of insertion points, it can
be shown that the computation changes by a total derivative
\refs{\Vpart,\ars}
\eqn\totald{\int_{\cal F} ~\partial \omega}
\noindent
where ${\cal F}$ is the appropriate fundamental domain of integration
for the computation.  Therefore, one must worry about contributions
arising at the boundary of ${\cal F}$ \amsgl.

\subsec{The Fundamental Domain}

What is the fundamental domain ${\cal F}$ for this computation?
At genus two, the Teichmuller space is given very explicitly
in terms of the Siegel upper half space of $2\times 2$ matrices:

$${\cal H}_{2} = \{\tau_{2 \times 2} :~~ \tau^{tr} ~=~
\tau,~~ {\rm Im} ~\tau ~>~ 0 \}$$

\noindent
$\tau$ is the period matrix of the genus two surface.
The modular group at genus two is $G = Sp(4,\IZ)$.
The moduli space can then be constructed by taking the quotient of
${\cal H}_{2}$ by $G$.
One must also remove the modular orbit of the diagonal matrices.

For our computation, on the other hand, we have twists $(1,1,f,g)$ about
the $(a_1,a_2,b_1,b_2)$ cycles of the surface.  Therefore, we need to
integrate the correlator of the picture changing operators over
${\cal F} = {\cal H}_{2}/{\tilde G}$, where $\tilde G$ is the
subgroup of $Sp(4,\IZ)$ which preserves the twist structure
$(1,1,f,g)$.

It is easy to see that the allowed matrices are the ones that act
on the homology $(a_1,a_2,b_1,b_2)$ like

\eqn\allmat{\pmatrix{ a & b & 0 & 0 \cr
                      c & d & 0 & 0 \cr
                      x & y & 1 & 0 \cr
                      z & w & 0 & 1 }}

\noindent
Denoting the $2\times 2$ blocks as
$\pmatrix { A & B \cr C & D }$
we must impose

\eqn\spcond{A^{tr} C = C^{tr} A,~~B^{tr} D = D^{tr}B,~~A^{tr}D - C^{tr}B = 1}

\noindent
which is just the requirement that \allmat\ is in $Sp(4,\IZ)$.
This further restricts the allowed matrices \allmat\ to be of the
form

\eqn\finalmat{\pmatrix{ 1 & 0 & 0 & 0 \cr
                        0 & 1 & 0 & 0 \cr
                        x & y & 1 & 0 \cr
                        y & w & 0 & 1}}

Now, if $\pmatrix{ A & B \cr C & D }$ acts on the homology, then
the action on the period matrix $\tau$ is given by $\pmatrix{ D & C \cr B &
A}$ -- in other words,

\eqn\tauact{\tau \rightarrow (D\tau ~+~ C) ~(B\tau ~ +~ A)^{-1}}

\noindent
So from the allowed actions on the homology \finalmat, we see that
the identifications to be made on the period matrices are

\eqn\periodid{\pmatrix{\tau_1 & \tau_{12} \cr
                       \tau_{12} & \tau_{2} }
\rightarrow \pmatrix{\tau_{1} + x & \tau_{12} + y \cr
                     \tau_{12} + y & \tau_{2} + w }}

\noindent
In addition, positivity of ${\rm Im} ~\tau$ requires that

\eqn\imtau{{\rm Im} ~\tau_{1,2} > 0,~~({\rm Im} ~\tau_{12})^{2} < {\rm Im}
~\tau_{1} ~{\rm Im} ~\tau_{2}}

The constraints \periodid\ and \imtau\ together yield the correct
fundamental domain ${\cal F} \subset {\cal H}_{2}$ for our computation.
$\tau_{1,2}$ live on strips with real part between
$(-1/2, 1/2)$ and positive imaginary part, while $\tau_{12}$ has
real part between $(-1/2, 1/2)$ and imaginary part bounded above
and below by the second inequality in \imtau.  Also, we must recall that
in describing the moduli space of Riemann surfaces in terms of ${\cal H}_{2}$,
we had to delete the modular orbit of diagonal matrices, yielding
an additional
boundary at $\tau_{12} ~\rightarrow ~ 0$.

\subsec{The Boundaries}

Now that we have determined ${\cal F}$, we can look for boundaries where
the total derivative
\totald\ might give a contribution after integration by parts.
There are in fact three boundaries in ${\cal F}$.
We will examine each of these boundaries in turn, and argue that
no boundary contribution
exists.

\noindent
1) $\tau_1$ or $\tau_2 ~\rightarrow~ i \infty$

\smallskip
\epsfbox{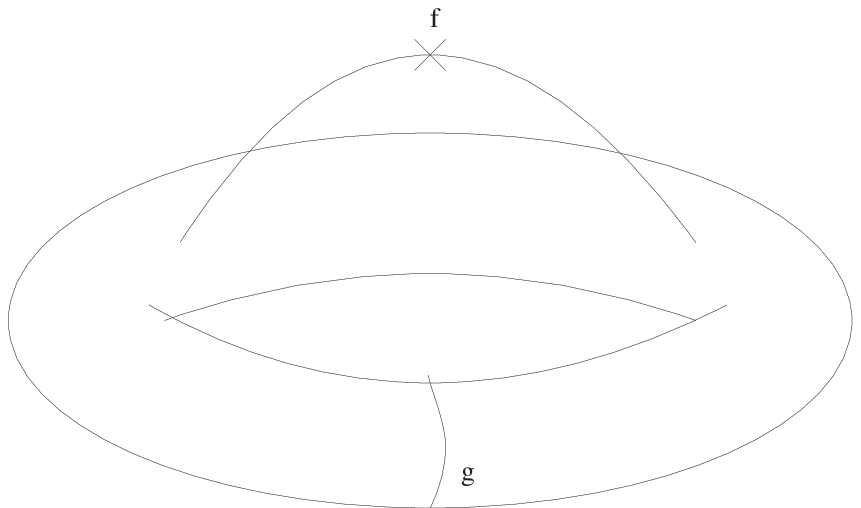}
\smallskip
\centerline{Figure 4: Picture of boundary 1).}
\smallskip

\noindent
In this limit, one of the handles degenerates to a semi-circle
glued on to the ``fat" handle at two points (i.e. a homology
cycle collapses).
It was argued in \amsgl\ that in such a limit, no boundary contribution
exists in theories without physical tachyons.  Our theory
has no physical tachyons, so we will receive no contribution from this
boundary.

\medskip
\noindent
2) $\tau_{12} ~\rightarrow ~0$

\smallskip
\epsfbox{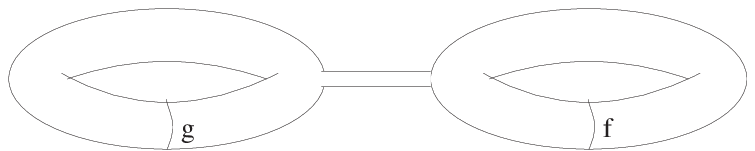}
\smallskip
\centerline{Figure 5: Picture of boundary 2).}
\smallskip

\noindent
In this limit, the genus two surface degenerates into two tori connected by
a very long, thin tube.
Only massless physical states propagate in this tube \amsgl,
and in this limit the
genus two vacuum amplitude is related to a sum of products of
one loop tadpoles for the massless states.

The relevant one loop tadpoles are computed on tori with twists
$(1,f)$ or $(1,g)$ around the $(a,b)$ cycles.  Now, the $f$ and $g$
twist alone preserve $d=4,~{\cal N}=2$ supersymmetry.  So, there are
no one loop tadpoles for states in the $f$ or $g$ twisted theory.
This implies that the genus two diagram vanishes in this limit.

\medskip
\noindent
3) ${\rm Im} ~\tau_{1,2} ~\rightarrow~0$ or $({\rm Im}~\tau_{12})^2 ~\rightarrow
~({\rm Im}~\tau_1) ~({\rm Im}~\tau_2)$

\noindent
To see the vanishing in this limit, we recall that the integrand for
the vacuum amplitude contains a factor of $e^{-S(X)}$, i.e. the action
for map from the genus two surface to spacetime.  The relevant maps
(given the $f$ and $g$ twists about the $b$ cycles of the surface)
wind around the $X_5$ and $X_6$ directions of spacetime.  This yields
a contribution to the action which goes like

\eqn\action{S \simeq {R^2\over \alpha^\prime} \{ {{{\rm Im} ~\tau_1 + {\rm Im} ~
\tau_2 }\over { {\rm Im} ~\tau_1 ~{\rm Im} ~\tau_2 ~-~({\rm Im} ~\tau_{12})^2
}} \}}

\noindent
where $R$ is the radius of the $X_5$ and $X_6$ circles
\refs{\polchinski,\mcclain}.
Now, positivity
of ${\rm Im} ~\tau$ comes to the rescue:

\noindent
$\bullet$ If ${\rm Im}~\tau_1 ~\rightarrow ~0$ at fixed ${\rm Im} ~\tau_2$,
then the second inequality in \imtau\ implies that $S \rightarrow \infty$.

\noindent
$\bullet$ If ${\rm Im}~ \tau_{1,2} ~\rightarrow ~0$, one can prove that
the denominator in \action\ vanishes as the square of the numerator (once
again using positivity of ${\rm Im} ~\tau$), so $S \rightarrow \infty$.

\noindent
$\bullet$ If ${\rm Im}~\tau_{1,2}$ are fixed and $({\rm Im}~\tau_{12})^2$
approaches ${\rm Im}~\tau_1 ~{\rm Im}~{\tau_2}$, it is obvious that the
action diverges.

\medskip
\noindent
The upshot is that the $e^{-S(X)}$ in the integrand
vanishes quickly enough at this boundary to rule out any
contributions.

\subsec{Cases with Shifts}

In additional to the amplitude $(1,1,f,g)$ which knows about
the supersymmetry breaking at genus two, there are other
genus 2 amplitudes with $(1,1,f,g)$ twists on the worldsheet
fermions around the $(a_1,a_2,b_1,b_2)$ cycles but with additional
shifts acting on the bosonic fields.  In fact we show in \S5\ that
this is (up to modular transformations) the full set of supersymmetry
breaking diagrams that we need to consider at genus two.

In each twist structure we will find that the spin-structure
dependent part of the vacuum amplitude vanishes.
This leaves the issue of possible boundary contributions.
After
summing over the various twist structures, we know the
genus two vacuum energy can be written as an integral over
${\cal M}_{2}$, the moduli space of genus two Riemann surfaces.
The possible boundary contributions (after we compactify ${\cal M}_{2}$)
will come from
boundaries of type 1) and 2) in \S4.2\
(where a handle collapses
or the surface degenerates into two surfaces of lower genus connected
by a long, thin tube).
Hence, if we can argue that with arbitrary twist and shift
structures on the $a,b$ cycles the vacuum amplitude vanishes
at boundaries of type 1) and 2), we will be done.

As we will discuss in \S5, up to additional shifts on various cycles
the possible structures (which break all of the supersymmetry)
are basically $(1,1,f,g)$, $(f,g,g,f)$
and $(f,fg,fg,f)$ (up to possible exchanges of the role of
$f$ and $g$).
Since we could use modular transformation to relate these to
$(1,1,f,g)$ twist structure on the fermions, the spin-structure
dependent piece of the amplitude vanishes in each of these cases.
In addition, each of these vanishes at boundaries of type 1)
because there are no physical tachyons.  This leaves the analysis
of boundary 2).

Any amplitude with $(1,1,f,g)$ twists on the fermions, regardless of
additional shifts, vanishes at boundary 2) because it can be
written as a product of tadpoles in the ${\cal N}=2$ supersymmetric
$f$ and $g$ orbifolds (as in \S4.2).  On the other hand, the amplitude with
twist $(f,g,g,f)$ would naively yield a product of one loop
tadpoles in a nonsupersymmetric theory.
However, it turns out that the state propagating on the
tube between the
first and second handle must be
a massive state because it must be twisted to be emitted
from the ``subtorus'' with $(f,g)$ twist on its $(a,b)$ cycles.
Since only massive states can run in the tube, there is no contribution
at the boundary of moduli space (where the tube becomes infinitely
long).  A similar discussion applies to the $(f,fg,fg,f)$ twist
structure with arbitrary shifts.

\newsec{Twists at Genus $h \geq 2$}

A priori on a genus $h$ Riemann surface, one needs to consider
any combination of twists on the various cycles $a_i, b_i$ for
$i=1,\cdots,h$ consistent with the relation

\eqn\relation{a_1 b_1 a_1^{-1}b_{1}^{-1}\cdots a_h b_h a_h^{-1}b_h^{-1} ~=~1}

\noindent
In this section, we will argue that in fact using modular transformations
one can greatly reduce the kinds of twist structures that one needs to
consider.

For our considerations, we do not need to worry about twists that
preserve some of the spacetime supersymmetry at genus $h$ (for instance,
twists only by $f$ around various cycles).  The real concern will be
sets of twists around different cycles which break the full spacetime
supersymmetry.  We will now show that, up to inducing shifts on the
worldsheet bosons around some cycles, one $\it only$ has to consider
$f$ and $g$ twists on the $b_{h-1}$ and $b_h$ cycles with no twists on any
other cycles.  Any twist which breaks all of the spacetime
supersymmetry can be brought to this canonical $(1,\cdots 1,f,g)$ form
by modular transformations.

Since in this section we will be ignoring the
possible shifts on bosons around various cycles (we're only interested
in the $f,g$ action on fermions), we can use relations like

\eqn\almrel{f^2 ~=~ g^2 ~= ~1,~~ fg = gf}

\noindent
which are true for the action on
fermions (but only true in the full model up to shifts in the space group).

\subsec{Genus $h=2$}

We will show that all twists of interest can be taken
to the $(1,\cdots,f,g)$ form in several steps.
First, consider genus two surfaces.
The modular group $Sp(4,\IZ)$ is generated by

\eqn\modgrpa{D_{a_1} = \pmatrix{1 & 0 & 0 & 0 \cr
                               0 & 1 & 0 & 0 \cr
                               1 & 0 & 1 & 0 \cr
                               0 & 0 & 0 & 1 }
,~~D_{a_2} = \pmatrix{1 & 0 & 0 & 0 \cr
                      0 & 1 & 0 & 0 \cr
                      0 & 0 & 1 & 0 \cr
                      0 & 1 & 0 & 1 }}
\eqn\modgrpb{D_{b_1} = \pmatrix{1 & 0 & 1 & 0 \cr
                                0 & 1 & 0 & 0 \cr
                                0 & 0 & 1 & 0 \cr
                                0 & 0 & 0 & 1 }
,~~D_{b_2} = \pmatrix{0 & 1 & 0 & 1 \cr
                      0 & 1 & 0 & 0 \cr
                      0 & 0 & 1 & 0 \cr
                      0 & 0 & 0 & 1 }}
\eqn\modgrpaa{D_{a_{1}^{-1} a_{2}} = \pmatrix{1 & 0 & 0 & 0 \cr
                                            0 & 1 & 0 & 0 \cr
                                            -1 & 1 & 1 & 0 \cr
                                             1 & -1 & 0 & 1 }}

\noindent
which are simply the Dehn twists about the various cycles of the genus
two surface, acting on the homology $(a_1,a_2,b_1,b_2)$.\foot{There
are also inhomogeneous terms that shift the characteristics of the
theta functions coming from the fermion determinants under such a
modular transformation;
these lead to a change of spin structure but do not change the orbifold
twist structure.}

We now consider genus $h=2$ twists which are not of the
canonical $(1,1,f,g)$ form but which break all the supersymmetry:

\noindent 1) First,
take the cases where no ``subtorus" has twists which break the full
supersymmetry (i.e., no $f,g$ twists on dual $(a,b)$ cycles).  Then,
by using $SL(2,\IZ) \subset Sp(4,\IZ)$
transformations which act on the $(a_1,b_1)$ and $(a_2,b_2)$ cycles,
one can arrange to have twists only on the $b$ cycles, so the twist
structure is $(1,1,*,*)$.  Then the only cases we need to worry about
are $(1,1,f,fg)$ and $(1,1,g,fg)$.  One can easily see that
$(1,1,f,fg)$ is mapped by
$D_{b_1}$ to $(f,1,f,fg)$ and then by $D_{a_1^{-1}a_2}$ to $(f,1,1,g)$.
That in turn is $SL(2,\IZ)$ equivalent to $(1,1,f,g)$.
A similar manipulation works for the $(1,1,g,fg)$ case.

\medskip
\noindent 2) Second, consider the case where there $\it are$ twists
on some ``subtorus'' that break the full supersymmetry.  Examples are
$(f,g,g,f)$ and $(f,fg,fg,f)$.  Now, for instance, $(f,g,g,f)$ can be
mapped by $D_{a_1^{-1}a_2}$ to $(f,g,f,g)$ which is equivalent
(using $SL(2,\IZ)$ transformations on both subtori) to $(1,1,f,g)$.
One can similarly reduce $(f,fg,fg,f)$ and other analogous structures
to the canonical form.  (Recall that in this discussion we are
ignoring extra bosonic shifts that make the twist structures considered
here consistent with \relation).

So, we find that $\it all$ supersymmetry breaking twists at genus
$h=2$ can be mapped by the modular group to $(1,1,f,g)$ (up to
shifts on worldsheet bosons).  This is important because our vanishing
at $h=2$ was for the spin structure dependent part of precisely
this twist structure, and is independent of any shifts on worldsheet
bosons.

\subsec{Genus $h>2$}

We now argue that at arbitrary genus, one can reduce all supersymmetry
breaking twist structures to $(1,\cdots,1,f,g)$ using modular
transformations.  We will need to use three important facts:

\noindent
1) Among the elements of $Sp(2h,\IZ)$ there are matrices that allow one
to permute the different ``subtori'' (sets of conjugate $a,b$ cycles)
of the genus $h$ surface.

\noindent
2) In order to satisfy \relation, there must exist an $\it even$ number
of ``subtori'' with twists on the $(a_i,b_i)$ cycles that break all
the supersymmetry.

\noindent
3) Using $Sp(4,\IZ) \subset Sp(2h,\IZ)$ one can map
\eqn\simplify{(1,1,f,f)~ \rightarrow ~(1,1,f,1)}
i.e. one can group like twists on neighboring $b$ cycles onto a
single $b$ cycle.

Putting together our $h=2$ result with facts 1)-3) above, we see
that at genus $h>2$ the only twist structure we need to consider is
$(1,1,\cdots,1,f,g)$.  To prove this, we simply work on genus 2
subsurfaces (using $Sp(4,\IZ)$ subgroups of the modular group)
to reduce everything to $f$ or $g$ twists on b cycles, and then
use 1) and 3) to simplify to a single $f$ and $g$ twist.

\newsec{Comments on Higher Loop Vanishing}

Once we have put the twists on our genus $h$ surface $\Sigma_h$ into
the canonical $(1,\cdots,1,f,g)$ form, we can provide a rough
physical argument for the vanishing.
This section is very heuristic;
it would be nice to make
these arguments more precise.

The argument involves supersymmetry.  One can think of
$\Sigma_h$ in terms of a genus $h-1$ surface $\Sigma_{h-1}$
(with a $g$ projection on one cycle) connected to an extra
handle (holding the $f$ projection) by a nondegenerate tube (on which
massive or massless string states may propagate).

\smallskip
\epsfbox{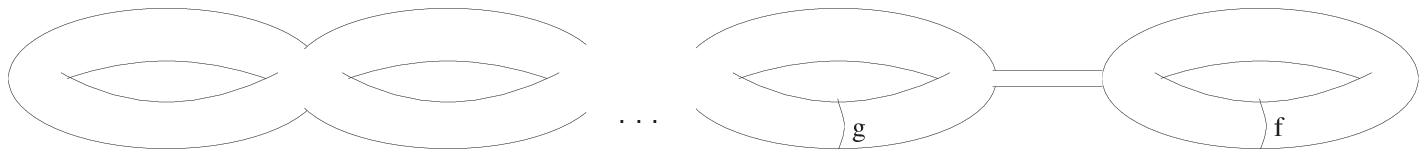}
\smallskip
\centerline{Figure 6: One sums over states in the channel
between the $f$ and $g$ twisted handles.}
\smallskip

\noindent
This suggests that one rewrite the diagram as

\eqn\factor{\Sigma_{\rm s} ~({\rm s ~tadpole~ on~ \Sigma_{h-1}})~ \times
e^{-M_{s}T}~ \times ({\rm s ~tadpole ~on~ f-projected ~handle})}

\noindent
where the sum runs over possible intermediate physical states
$s$ of mass $M_s$, and the tube has length $T$.
In this way of thinking about it, the diagram vanishes because
even for massive string states,
the tadpoles at genus $h-1$ in the $g$ projected theory and at genus one
in the $f$ projected theory should vanish (as those theories are
both 4d ${\cal N}=2$ supersymmetric).

\subsec{Toward a (perturbative) symmetry argument}

The worldsheet arguments for the perturbative vanishing of the cosmological
constant in supersymmetric string theories
used contour deformations of the spacetime supercurrents crucially
\refs{\martinec,\Vpart,\amscat}.
Thus one could see the constraint of spacetime supersymmetry
through worldsheet current algebra.

In our theory, of course, the spacetime supercurrents are projected
out, which is to say that they have monodromy around the twist
fields in the orbifold.  On the Riemann surface with a given
twist structure (as in Figure 6), these operators pick up
phases upon traversing the cuts in the diagram.

Let us consider the argument of \martinec\ in this context.
One splits open one handle of the surface (say the handle
with the $f$-cut), and rewrites the
propagating state $V$  as $\oint_{a_f} S_{+-} V^\prime$ for some operator
$V^\prime$.  (So in particular $V^\prime$ describes
a boson if the original state was fermionic in spacetime and vice
versa).  Here $S_{+-}$ refers to a would-be spacetime supercurrent
with eigenvalues $+1$ under $f$ and $-1$ under $g$.  The cycle
$a_f$ is the $a$-cycle on the handle with the $f$ cut.
Without the $g$ cut, one can deform the contour integral around
the rest of the Riemann surface and turn the fermion loop
into a boson loop (or vice versa) with a cancelling sign.
With the $g$-cut, however,
one is left with a remainder contribution of the form
\eqn\remainder{
2 \oint_{a_f}dx \oint_{a_g}dy ~\langle S_{+-}(x) S_{+-}(y) \rangle}

The direct calculation of this contribution could be done
at genus 2 in a similar way to the partition function calculation
we presented in the previous sections (using the
correlation functions of \Vcorr).  As before it would
be hard to then generalize this computation to higher genus
precisely, due to our lack of explicit understanding of the
moduli space (and the problem of choosing a consistent
gauge slice for the worldsheet gravitino).  It would be nice to
understand if there is some simple topological reason that this
remainder must vanish at arbitrary genus.

\newsec{Relation to AdS/CFT Correspondence}

There has been a great deal of recent work on the fascinating
conjectures relating conformal field theories in various dimensions
to string theory in Anti de Sitter backgrounds \refs{\juan,\ads}.\foot{
Indeed, the subject has been as popular as the
Chicago Bulls.}
It was argued in \ks\ that certain nonsupersymmetric instances of
this correspondence could lead to the discovery of nonsupersymmetric
string backgrounds with vanishing cosmological constant.
The predictions of new fixed lines at the level of the leading
large-$N$ theory based on the correspondence \ks\ were verified
directly through remarkable cancellations in perturbative diagrams
\bkv\ in a host of models that could be constructed
quite systematically \lnv.
Here, we review and elaborate on the idea of extending
predictions of
the correspondence to finite-$N$ fixed lines, and explain how the
class of orbifolds we have discussed in this paper may be realizations.

The correspondences between 4d CFTs and string backgrounds have
(in the 't Hooft limit)

\eqn\mapone{{\alpha^\prime \over R^2} ~{\rm expansion} ~\rightarrow
~{\rm expansion ~in}~(g_{YM}^{2}N)^{-{1\over 2}}}
\eqn\maptwo{g_{st} ~{\rm expansion} \rightarrow ~{\rm expansion ~in}~
g_{YM}^{2}~=~{1\over N}}

\noindent
In cases where one has a $\it nonsupersymmetric$ fixed line
(to all orders in ${1\over N}$ as well as $g_{YM}^{2}N$) realized
on branes in string theory, we would
obtain by this correspondence a stable nonsupersymmetric string
vacuum which exists at arbitrary values of the coupling $g_{st}$.

The equation of motion for the dilaton $g_{st} = e^{\phi}$
is

\eqn\dileq{(-g)^{1\over 2} \partial_{\mu} (\sqrt{(-g)}
g^{\mu \nu} \partial_{\nu}
\phi) ~=~-{{\partial V}\over {\partial \phi}}}

\noindent
where $V(\phi)$ is the dilaton potential and $g$ is the
$AdS$ metric.
Thus one finds that, according to the AdS/CFT correspondence,
a conformal fixed point in the dual field theory implies a minimum
of the bulk vacuum energy with respect to the $g_{st}$.
Consequently, stability of the
spacetime background for arbitrary dilaton VEV $\langle \phi \rangle$
(which is implied by the existence of a nonsusy fixed line as above)
would imply that there is no $g_{st}$ dependent vacuum energy.

Non-supersymmetric theories with vanishing $\beta$-function
at leading order in $1/N$ but nonzero $\beta$-functions at
subleading orders \refs{\ks,\lnv,\bkv} should
provide a concrete testing ground for
ideas relating the holographic principle \holography\ to
the cosmological constant \tom.  In particular, one would
generically expect perturbative contributions to the cosmological
constant (dilaton tadpole), which are related (via a possibly nontrivial map)
to beta functions in the boundary field theory.  In perturbative
string theory these contributions come in generically at the order of
the supersymmetry
breaking scale, which is the string scale in
these models.
In general, the AdS/CFT correspondence relates perturbative
string corrections to $1/N$ corrections in the boundary
QFT.  Therefore the perturbative string corrections should
be encoded in the boundary theory in a way consistent with
the holographic reduction in the number of degrees of freedom.

Dualities between field theory fixed lines
(which exist to all orders in $1/N$) and nonsupersymmetric
string backgrounds would have different consequences in
the different $AdS_{d}$ dualities.
In the orbifolds $AdS_{5} \times ({S^{5}/\Gamma})$, the duality
would
imply that the effective $\it ten-dimensional$ cosmological constant
vanishes.  In the large $g_{YM}^{2}N$ limit, we expect from
\mapone\ that the $AdS$ and orbifolded sphere also each become flat.
However, in this limit the spacetime theory regains supersymmetry
away from the fixed loci of $\Gamma$, so this does not yield a
nonsupersymmetric theory in the bulk of spacetime.

However, there are cases where the large $N$ limit $\it could$ yield
a nonsupersymmetric theory in bulk with vanishing cosmological constant.
Consider for instance dualities between IIB strings
on $AdS_{2}\times S^{2}\times
({T^{6}/\Gamma})$
and conformally invariant quantum mechanical systems (some supersymmetric
instances of such dualities were conjectured in \juan).
In these cases, going through the analogous arguments we would be
talking about the effective $\it four-dimensional$ cosmological constant.
In the large $N$ limit, $AdS_{2}\times S^{2} ~\rightarrow~ \IR^{4}$
while the size of the $T^6/\Gamma$ remains $\it fixed$ (it does
not decompactify).  Thus, if we break supersymmetry on the internal
space we might be able to find examples of the AdS/CFT correspondence
which predict vanishing 4d cosmological constant in a bulk nonsupersymmetric
theory.  This provides a strong motivation for understanding
conformally invariant quantum mechanical systems with ``fixed lines''
(corresponding to the spacetime $g_{st}$).
In particular, at least naively a quantum mechanical model which is
classically conformal will not develop a $\beta$-function since
there are no ultraviolet divergences (though one may need to worry
about IR problems).

We have two comments about trying to find models in this way via
the AdS/CFT correspondence:

\noindent
1) The $AdS_2 \times S^2$ geometries of interest arise as the
near horizon limits of Reissner-Nordstrom black holes.  In
nonsupersymmetric situations, where $\pi_1$ of the
compactification is typically small, it can be very difficult to find
stable black holes of this sort by wrapping branes.  In part
this is because
one often finds a 4d effective Lagrangian of the form

\eqn\lagr{{\cal L} ~=~\int~d^{4}x~\phi F_{\mu\nu}F^{\mu\nu} ~+~
\partial^{\mu}\phi \partial_{\mu}\phi ~+ \cdots}

\noindent
where $F$ is the field strength for the $U(1)$ gauge field under
which the Reissner-Nordstrom black hole carries charge, and
$\phi$ is some scalar field.  Then, the equation of motion for
$\phi$ becomes

\eqn\eom{\partial^{\mu}\partial_{\mu}\phi~\sim~ F_{\mu\nu}F^{\mu\nu}}

\noindent
and this forces $\phi$ to have a nontrivial profile in the
black hole solution which breaks the $AdS$ isometry.

In order to get around problems of stability and of the existence
of scalars with linear couplings to $F^{2}$, it is useful to start
with models containing very few scalars.  Asymmetric orbifolds are
one natural source of such models.  Starting
with configurations of wrapped branes invariant under the orbifold
group, one can obtain Reissner-Nordstrom black holes in
asymmetric orbifolds such as the one we have studied here.  One can then
predict vanishing cosmological constant based on the conformally
invariant family of quantum mechanical systems living on the
boundary of the near-horizon geometry, as in the argument above.
It is intriguing that this rather indirect argument relates the
problem of fixing moduli to the cosmological constant problem.
It would be nice to understand the constraints more systematically.

\noindent
2) The orbifold we've been discussing not only has
\eqn\orderby{\Lambda_{1-loop} ~=~\Lambda_{2-loop} ~=~\cdots~=~0}
but also has
\eqn\idenze{\Lambda_{1-loop} ~=~\int~d^{2}m_i~0,~~\Lambda_{2-loop} ~=~
\int~d^{6}m_{i}~0,~\cdots}
That is, the vacuum amplitudes vanish $\it point~by~point$ on the moduli
space of Riemann surfaces(within a particular twist structure: it will
vanish point-by-point on all twist structures if one also acts on the gauge-fixing
choice with the relevant modular transformations).
This vanishing integrand reflects the
exceptionally simple spectrum of our theories (bose-fermi degeneracy, etc.).

In more general examples that might come out of nonsupersymmetric
versions of the AdS/CFT correspondence as above, we would expect
\orderby\ to hold (since the conformal quantum mechanics has a fixed
line, and the dilaton VEV is arbitrary).  However, there is no
reason to expect \idenze\ to hold in general examples.  It would be
nice to find an example where e.g. $\Lambda_{1-loop}$
vanishes but not point-by-point (as in Atkin-Lehner symmetry \moore).

It would be very interesting to find similar models with a more
realistic low energy spectrum.
In addition, we could potentially find non-supersymmetric
models in 4d with no dilaton by finding 3d string models satisfying
our conditions and taking $g_{st}\to\infty$.

\medskip
\centerline{\bf{Acknowledgements}}

We would like to thank Z. Kakushadze, G. Moore and H. Tye for
several very important discussions.
We also gratefully acknowledge very helpful
conversations with N. Arkani-Hamed, R. Dijkgraaf,
M. Dine, L. Dixon, J. Harvey, M. Peskin, A. Rajaraman, C. Vafa and
H. Verlinde.
S.K. is supported by NSF grant PHY-94-07194, by DOE contract
DE-AC03-76SF0098, by an A.P. Sloan Foundation Fellowship
and by a DOE OJI
Award.  J.K. is supported by NSF grant PHY-9219345 and by
the Department of Defense NDSEG Fellowship program.  E.S. is
supported by DOE contract DE-AC03-76SF00515. The authors also
thank the Institute for Theoretical Physics at Santa Barbara,
the Harvard University Physics department, and the
Amsterdam Summer Workshop on String Theory
for hospitality while portions of this work were completed,
and the participants of the associated workshops for discussions.

\listrefs
\end